\documentclass{article}
\usepackage{amsmath,amssymb}
\usepackage{graphicx}%
\newcommand{\Fig}[2]{%
\begin{center}
\parbox{8cm}{%
\refstepcounter{figure}\includegraphics[width=8cm]{#1} \noindent Figure \thefigure:\quad
#2}\end{center}}
\begin{document}
%%%%%%%%%Authors definitions%%%%%%%%%%%%%%%%
\begin{center}
{\bf \Large A macroscopic view to the standard cosmological model} \\[12pt]
Yu.G. Ignat'ev, D.Yu. Ignatyev and A.R. Samigullina\\
N.I. Lobachevsky Institute of Mathematics and Mechanics, Kazan Federal University, \\ Kremleovskaya str., 35, Kazan, 420008, Russia
\end{center}

\begin{abstract}
The averaging of the dynamic functions of the standard cosmological model (SCM) at its early stage with the dominance of the scalar field is carried out. It is shown that microscopic oscillations with Compton period provide the major contribution to the macroscopic energy density of the scalar field at large times at this stage. In this case, the effective equation of state of the scalar field oscillates between the inflationary and the extremely rigid ones, while the macroscopic equation of state is non-relativistic.

{\bf keyword} standard cosmological model, macroscopic averaging, fast oscillations, numerical simulation, numerical gravitation.
\end{abstract}

\section*{Introduction}
We investigate the standard cosmological model, based on classic massive scalar field at the early cosmological stage when scalar field dominates over other matter types (see eg.  \cite{Gorb})\footnote{
This work was funded by the subsidy allocated to Kazan Federal University for the state assignment in the sphere of scientific activities.}. There are numerous papers devoted to this problem, particularly, the problems of the qualitative theory for the Standard   Cosmological Model (SCM) were considered in papers \cite{Belinsky} -- \cite{Mex2}. In \cite{Ignat16-1} -- \cite{Ignat17-1} the asymptotic properties of the standard cosmological model, based on classical massive scalar field, were investigated by means of combined application of methods of the qualitative theory of ordinary differential equations and their numerical integration. In particular, it was shown, that the system of the Einstein -- Klein -- Gordon equations for the homogenous space-flat cosmological model has a singular point corresponding to zero values of the potential of the scalar field and its de\-ri\-vative, and in this case, the singular point can be an attractive center or an attractive focus or a saddle depending on a value of the cosmological constant. At $\Lambda<4/3m^2$ this singular point is an attractive center, while at $\Lambda=0$ it is an attractive focus\footnote{Here $m$ is a mass of scalar field's quanta in the Planck units  $G=c=\hbar =1$; this system is used in this paper.}. In addition, it was revealed that when approaching to this singular point, invariant cosmological acceleration takes an invariant cha\-ra\-cter with an average value, corresponding to non - rela\-ti\-vis\-tic equation of state. As a later research \cite{Ignat_Agafon17} has revealed, in the case of zero cosmological term, scalar field's oscillations and oscillations of its de\-ri\-va\-tive, decay with time, but this process lasts for a significantly long time, up to  $10^{7} $ Compton times. Here the phase trajectory of a dynamic system in the plane
$(\Phi ,\dot{\Phi })$ almost coincides with an ideal circle, which radius slowly decreases. This suggests that at suf\-fi\-ciently late times of the early evolution of the Universe\footnote{Speaking about large times of the early evolution we imply the circumstance that scalar field remains a sole type of matter in such times.} the main contribution in the macroscopic energy density of a scalar field is done by microscopic oscillations. In this article this concept is verified using numerical simulation methods, developed in Authors' works \cite{Ignat_Sam}.

\section{Main Relations of the Standard Cosmological Model}
As is well known, for the space-flat model
\begin{equation} \label{GrindEQ__4_}
ds^{2} =dt^{2} -a^{2} (t)(dx^{2} +dy^{2} +dz^{2} )
\end{equation}
the system of Einstein and Klein-Gordon equations consists of the unique nontrivial Einstein equation  ($\dot{f}\equiv df/dt$.):
\begin{eqnarray} \label{GrindEQ__5_}
3\frac{\dot{a}^{2} }{a^{2} } =\dot{\Phi }^{2} +m^{2} \Phi ^{2} \Rightarrow H^{2}=\frac{1}{3} \left(\dot{\Phi }^{2} +m^{2} \Phi ^{2} \right),
\end{eqnarray}
--  and the equation of the classical massive scalar field:
\begin{equation} \label{GrindEQ__7_}
\ddot{\Phi }+3H \dot{\Phi }+m_{}^{2} \Phi =0,
\end{equation}
where $H(t)$ is the Hubble constant $\Lambda =\ln (a);{\rm \; \; \; }\frac{\dot{a}}{a} \equiv \dot{\Lambda }=H(t).$
Tensor of the scalar field's energy -- momentum has a structure of energy -- momentum of the ideal isotropic flux with the following energy density and pressure:
\begin{equation} \label{GrindEQ__8_}
\varepsilon =\frac{1}{8\pi } \left(\dot{\Phi }^{2} +m^{2} \Phi ^{2} \right);\; p=\frac{1}{8\pi } \left(\dot{\Phi }^{2} -m^{2} \Phi ^{2} \right).
\end{equation}
Let us note the useful relation for definition the invariant cosmological acceleration$\Omega (t)$:
\begin{equation} \label{GrindEQ__9_}
{\rm \; }\Omega (t)=\frac{a\ddot{a}}{\dot{a}^{2} } \equiv 1+\frac{\dot{H}}{H^{2} } \equiv -\frac{1}{2}(1+3\varkappa),
\end{equation}
where $\displaystyle\varkappa =\frac{p}{\varepsilon}.$ is an effective barotrope coefficient.

\section{Reducing the System of Equations to Canonical Form}
Making use of the possibility to express the Hubble constant from the Einstein equation(\ref{GrindEQ__5_}) through functions $\Phi ,{\rm \; }\dot{\Phi }$, we proceed to the dimensionless Compton time:
\[mt=\tau ;\quad (m\rlap{$/$}\equiv 0)\]
and carry out the standard substitution of variables $\Phi '=Z(t)$. This allows us to reduce the system of field equations (\ref{GrindEQ__5_}), (\ref{GrindEQ__7_}) to the form of canonical autonomous system of ordinary differential equations in a 3-dimensional phase space $\{ \Lambda ,\Phi ,Z\} $:
\begin{equation} \label{GrindEQ__11_}
\Lambda '\equiv h=\sqrt{\frac{1}{3} \left(Z^{2} +\Phi ^{2} \right)} ;
\end{equation}
\begin{equation} \label{GrindEQ__12_}
\begin{array}{l} {{\rm \; }\Phi '=Z} \end{array} ;
\end{equation}
\begin{equation} \label{GrindEQ__13_}
Z'=-\sqrt{3} \sqrt{Z^{2} +\Phi ^{2} } Z-\Phi  ,
\end{equation}
where it is $f'\equiv {df\mathord{\left/ {\vphantom {df d\tau }} \right. \kern-\nulldelimiterspace} d\tau } $. Here it is:
\begin{equation} \label{GrindEQ__14_}
H=m\frac{a'}{a} \equiv mh;{\rm \; \; \; }\Omega =\frac{aa''}{a'^{2} } \equiv 1+\frac{h'}{h^{2} } .
\end{equation}
Let us also notice that system (\ref{GrindEQ__11_}) -- (\ref{GrindEQ__13_}) has an autonomous subsystem  (\ref{GrindEQ__12_}) - (\ref{GrindEQ__13_}) in the space $\{ \Phi ,Z\} $ \cite{Ignat16-1}. This subsystem was being investigated in papers \cite{Ignat16-1} -- \cite{Ignat17-1}.  In these works it was shown, that the system has only one attractive focus
\begin{equation} \label{GrindEQ__15_}
M0:{\rm \; \; \; (}\Phi _{\infty } =0;Z_{\infty } =0),
\end{equation}
onto which the phase trajectory is wound at $\tau \to \infty $. From the results of the paper \cite{Ignat_Agafon17} it follows, that  the phase trajectory at  $\tau \gg 1$ is nearly an ideal circle with slowly decreasing radius:
\begin{eqnarray}\label{GrindEQ__16_}
\Phi \simeq \Phi _{0}(\tau)\sin \tau ,& (\tau {\rm \gg }1); \\
\Phi ^{2} (\tau )+\Phi '^{2} (\tau )\approx \Phi _{0}^{2} (\tau ) & \displaystyle
\simeq \frac{1}{3\tau ^{2} }.
\end{eqnarray}
\section{The Time Averaging of the Oscillating Dynamic Variables}
Since at sufficiently large times of the early evolution $\tau \gg 1$ the dynamic system, which is described by equations (\ref{GrindEQ__11_})--(\ref{GrindEQ__13_}), performs fast oscillations with slowly changing amplitude, then following the standard procedure of macroscopic description of such a system, we should apply averaging of the fluctuating dynamic variables over time according to the following rule:
\begin{equation}\label{GrindEQ__17_}
\overline{\psi }(\tau )=\frac{1}{\Delta \tau } \int\limits_{\tau }^{\tau +\Delta \tau }\psi (\tau ')d\tau ',
\end{equation}
where $1\ll \Delta \tau \ll \tau $ is an averaging interval. Then oscillation (fluctuation) of the value $\psi (\tau )$ is found using formula:

\begin{equation} \label{GrindEQ__18_}
\Delta \psi (\tau )=\psi (\tau )-\overline{\psi }(\tau ),
\end{equation}
so that

\begin{equation} \label{GrindEQ__19_}
\overline{\Delta \psi }(\tau )\equiv 0.
\end{equation}
The following mean-square fluctuations are also important for us:

\begin{equation} \label{GrindEQ__20_}
\overline{\Delta \psi ^{2} }(\tau )=\overline{\left(\psi (\tau )-\overline{\psi (\tau )}\right)^{2} },
\end{equation}
\Fig{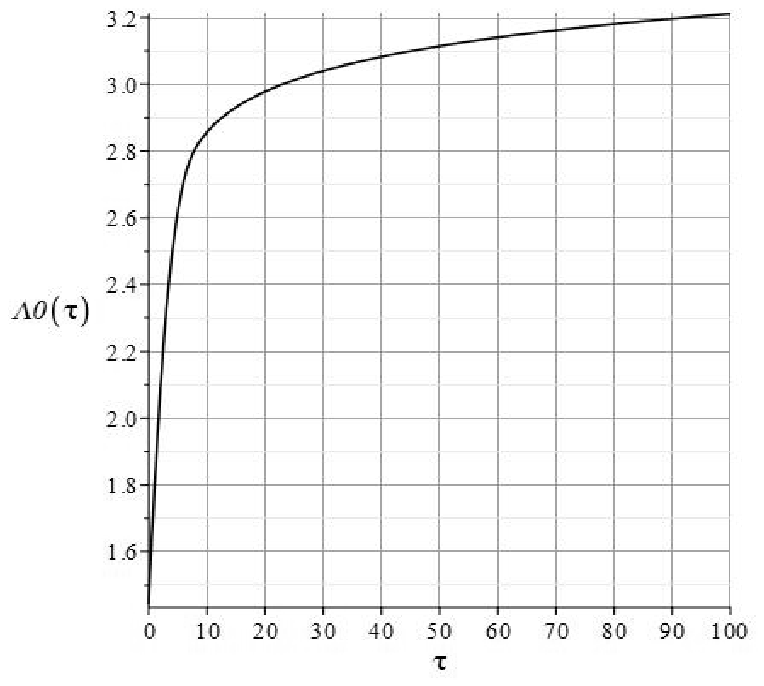}{Evolution of the average value $\Lambda _{0} (\tau )$ \label{img-2}}
\Fig{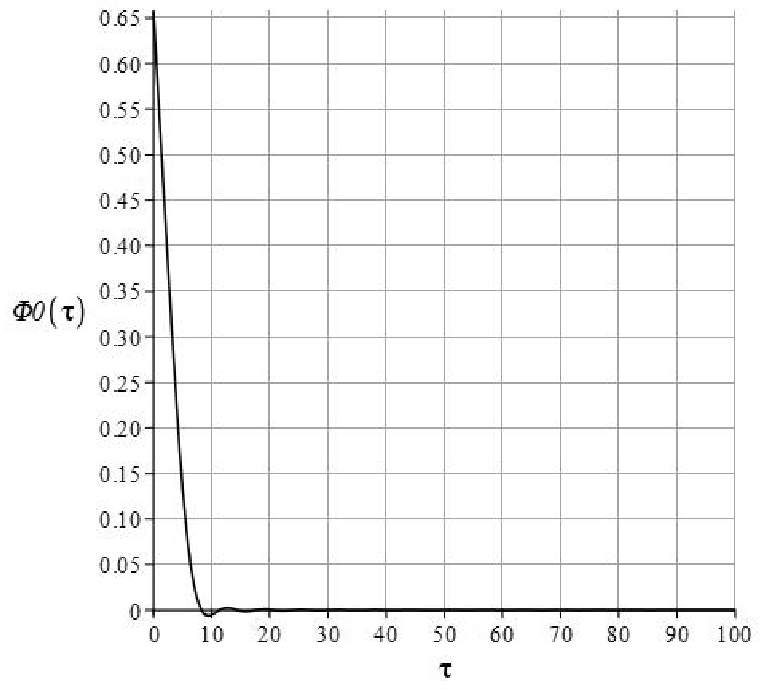}{Evolution of the average value $\Phi _{0} (\tau )$ \label{img-3}}

According to (\ref{GrindEQ__18_})--(\ref{GrindEQ__20_}) the following identity law takes place:

\begin{equation} \label{GrindEQ__21_}
\overline{\psi ^{2} (\tau )}\equiv \overline{\psi (\tau )}^{2} +\overline{\Delta \psi ^{2} (\tau )}.
\end{equation}
let us calculate firstly the mean values of the dynamic variables in correspondence with these definitions:
$$\Lambda _{0} (\tau )=\overline{\Lambda (\tau )};\quad \Phi _{0} (\tau )=\overline{\Phi (\tau )};\quad  Z_{0} (\tau )=\overline{Z(\tau )},$$
and then the deviations from the averages:
\begin{eqnarray}
\Delta \Lambda (\tau )=\Lambda (\tau )-\Lambda _{0} (\tau );\nonumber\\
\Delta \Phi (\tau )=\Phi (\tau )-\Phi _{0} (\tau );\nonumber\\
\Delta Z(\tau )=Z(\tau )-Z_{0} (\tau )\nonumber
\end{eqnarray}
and finally, their mean-square fluctuations:
$$\overline{\Delta \Lambda ^{2} }(\tau );\quad \overline{\Delta \Phi ^{2} }(\tau );\quad \overline{\Delta Z^{2} }(\tau ).$$

Since the system of differential equations (\ref{GrindEQ__11_}) -- (\ref{GrindEQ__13_}) is a sufficiently nonlinear one, let us apply numerical methods of integration from the Author's package DifEqTools \cite{Ignat_Sam},  which was developed speci\-fi\-cally for investigations of the nonlinear dynamic systems of any dimension in a system of computer maths Maple 18. For the solution of the differential equations we will use a specific Runge-Kutta inte\-g\-ration method of improved accuracy of 7-8 orders, while for inte\-g\-ration and differentiation we will use numerical methods of DifEqTools. Fig.\ref{img-2} -- \ref{img-5} illustrate certain inte\-g\-ration results.
\Fig{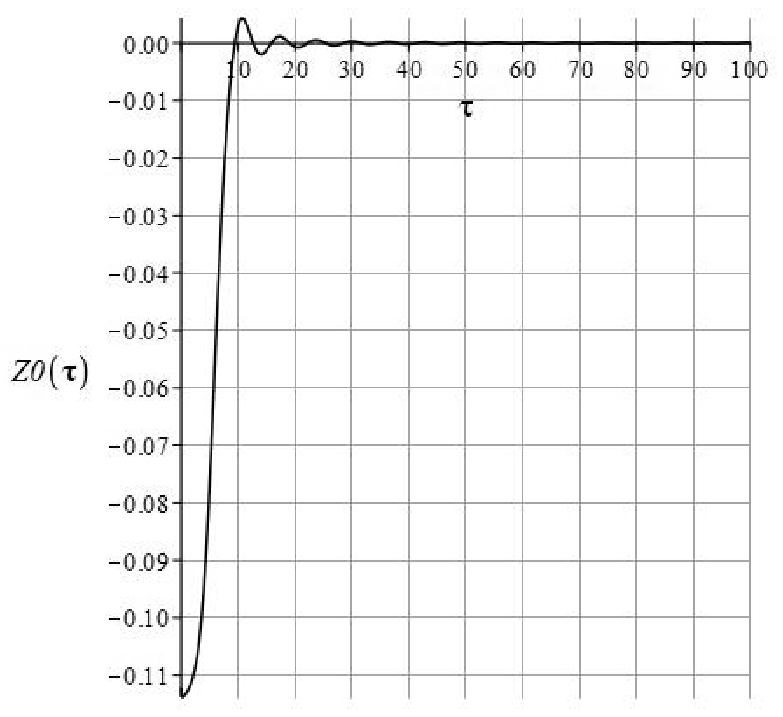}{Evolution of the average value $Z_{0} (\tau )$\label{img-4}}
Fig. \ref{img-6} illustrates the process of oscillation amplitude's decay while graph on Fig. \ref{img-7} clearly demonstrates a strict conservation of the oscillation period. This graph shows displays time interval containing 5 oscillation periods, each of which is equal to $2\pi $.\footnote{ let us remind, that we have chosen the Compton time scale $t_{c} =m^{-1} $ .}
\Fig{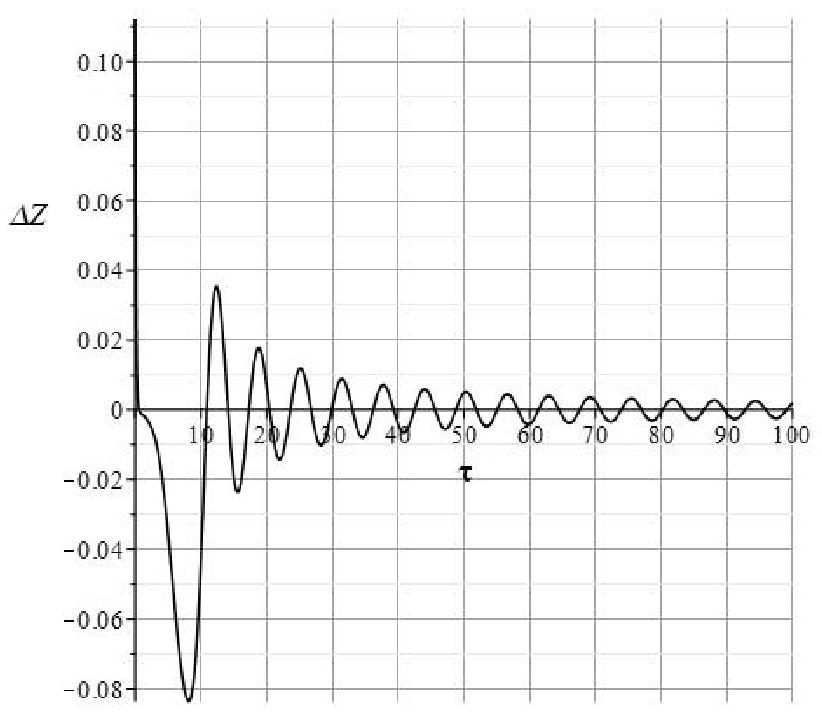}{Evolution of the oscillatons $\Delta Z(\tau )$\label{img-5}}
Fig. \ref{img-6}-\ref{img-7} display graphs of time evolution of the potential's oscillations $\Delta \Phi (\tau )$ in small and large scales.
\Fig{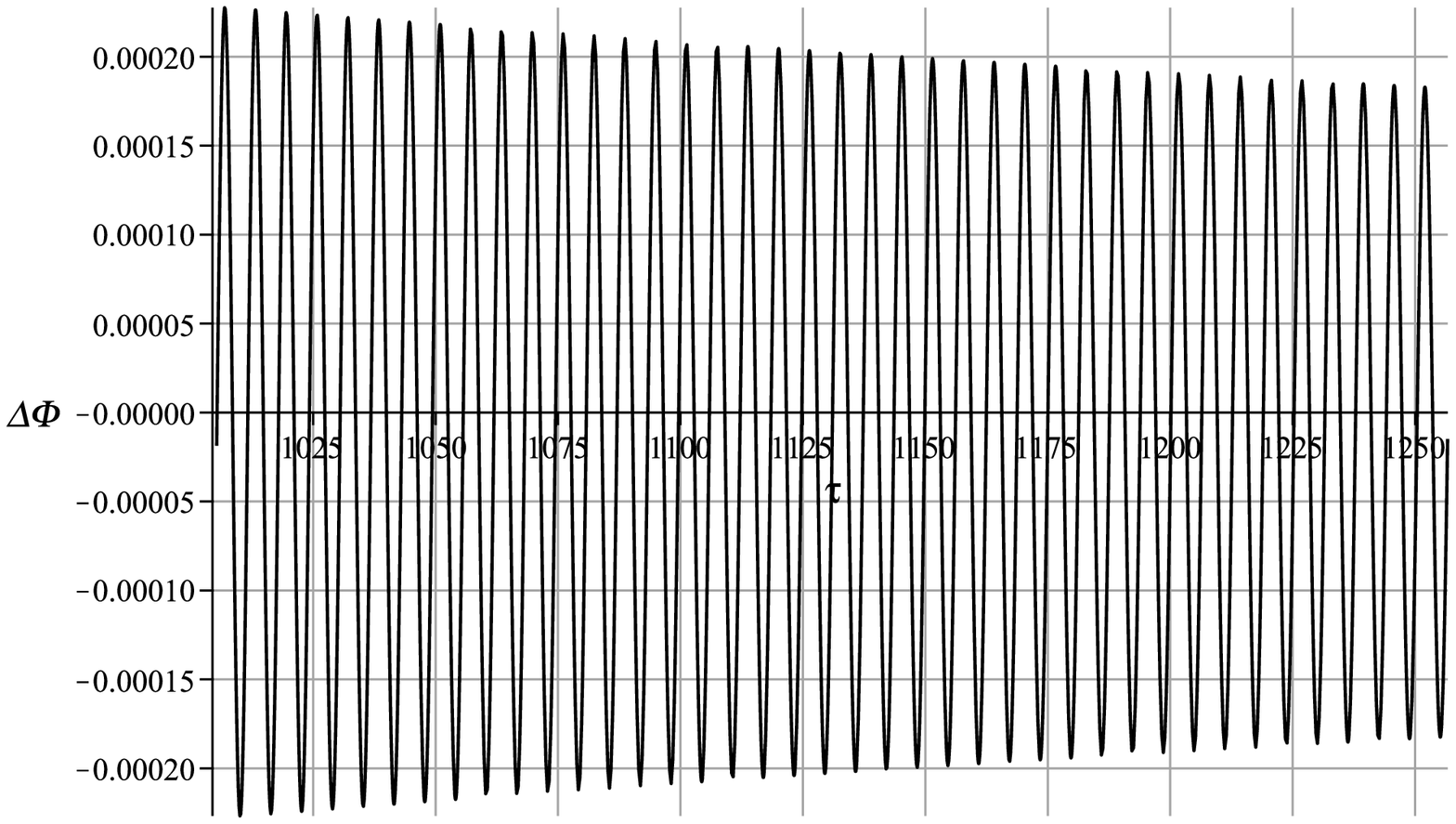}{Evolution of the oscillation $\Delta \Phi (\tau )$ in a small scale\label{img-6}}

\section{Time Averaging of Physical Properties}
Let us now proceed to calculation of the average physical scalar properties of the cosmological model: energy density,$\varepsilon (\tau )$, pressure, $p(\tau )$, the Hubble constant, $H(\tau )$, and related square fluctuations. According to formulas (\ref{GrindEQ__8_}) and (\ref{GrindEQ__21_}) we find:
\begin{equation} \label{GrindEQ__22_}
\overline{\varepsilon }=\varepsilon _{0} (\Phi _{0} ,Z_{0} )+\Delta \varepsilon ;{\rm \; \; }\overline{p}=p_{0} (\Phi _{0} ,Z_{0} )+\Delta p,
\end{equation}
where:
\begin{equation} \label{GrindEQ__23_}
\Delta \varepsilon =\frac{1}{8\pi } \left(\overline{\Delta Z^{2} }+\overline{\Delta \Phi ^{2} }\right);{\rm \; \; }\Delta p=\frac{1}{8\pi } \left(\overline{\Delta Z^{2} }-\overline{\Delta \Phi ^{2} }\right).
\end{equation}
Further, according to (\ref{GrindEQ__5_}) and (\ref{GrindEQ__14_}) we find for the average value of the Hubble constant
\begin{equation} \label{GrindEQ__24_}
\overline{h^{2} (\tau )}=\frac{8\pi }{3} \overline{\varepsilon }\Rightarrow \sqrt{\overline{h^{2} (\tau )}} =\sqrt{\frac{8\pi }{3} \overline{\varepsilon }} .
\end{equation}
First of all, Fig. \ref{img-8} shows the evolution of the energy density $\varepsilon _{0} (\Phi _{0} ,Z_{0} )$, calculated using formula (\ref{GrindEQ__8_}) with respect to average values of the scalar potential and its derivative. Secondly, it shows the evolution of the average energy density of the scalar field$\overline{\varepsilon (\tau )}$, calculated according to(\ref{GrindEQ__22_}).  Fig. \ref{img-9} shows similar graphs for the scalar field's pressure.

\Fig{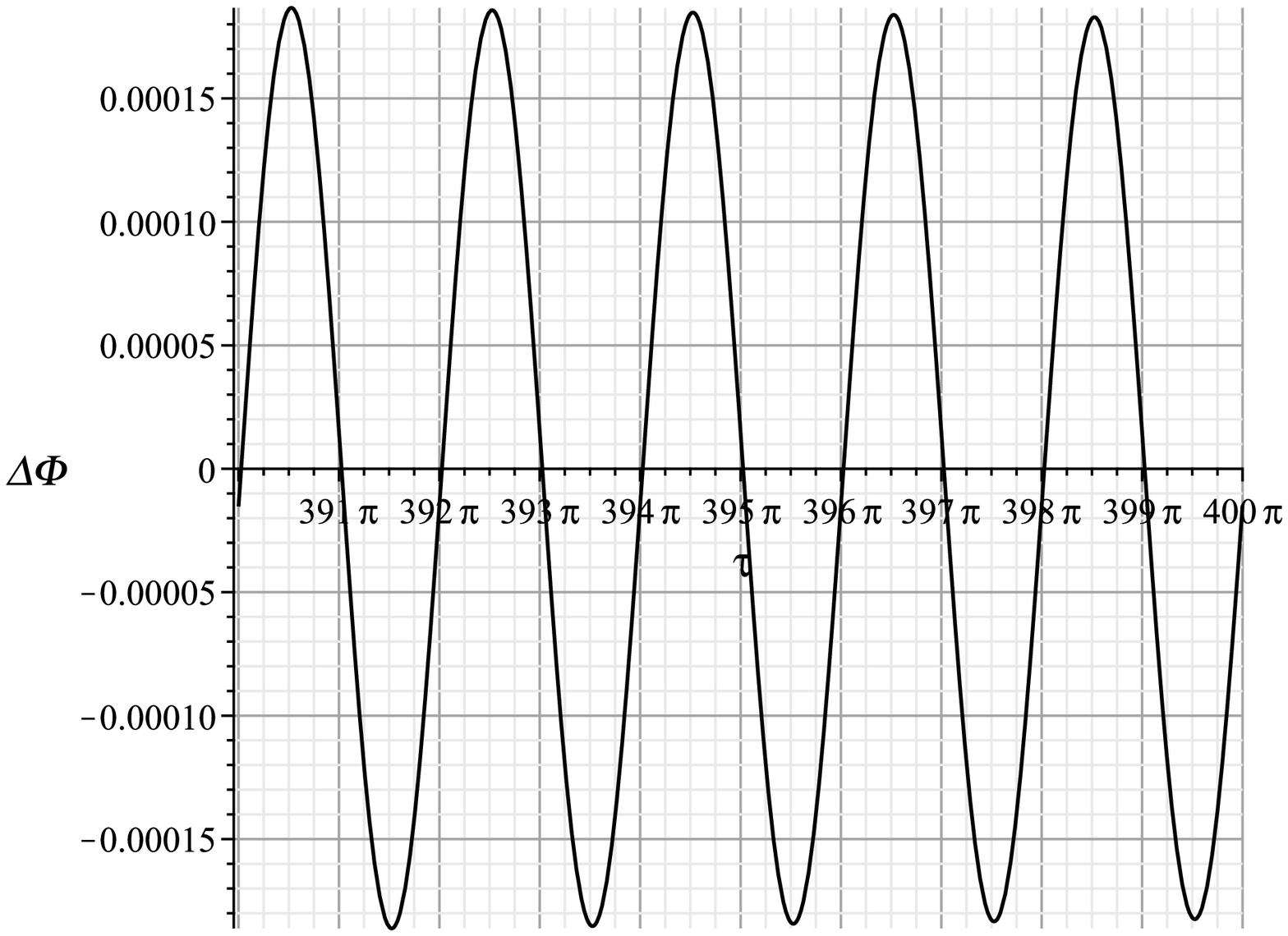}{Evolution of the oscillation $\Delta \Phi (\tau )$ in a large scale\label{img-7}}
As the results of numerical simulation illustrate, the scalar field's pressure, in contrast to energy density, oscillates around zero value at late times. Fig. \ref{img-10} represents evolution of the macroscopic value of the Hubble constant.
\Fig{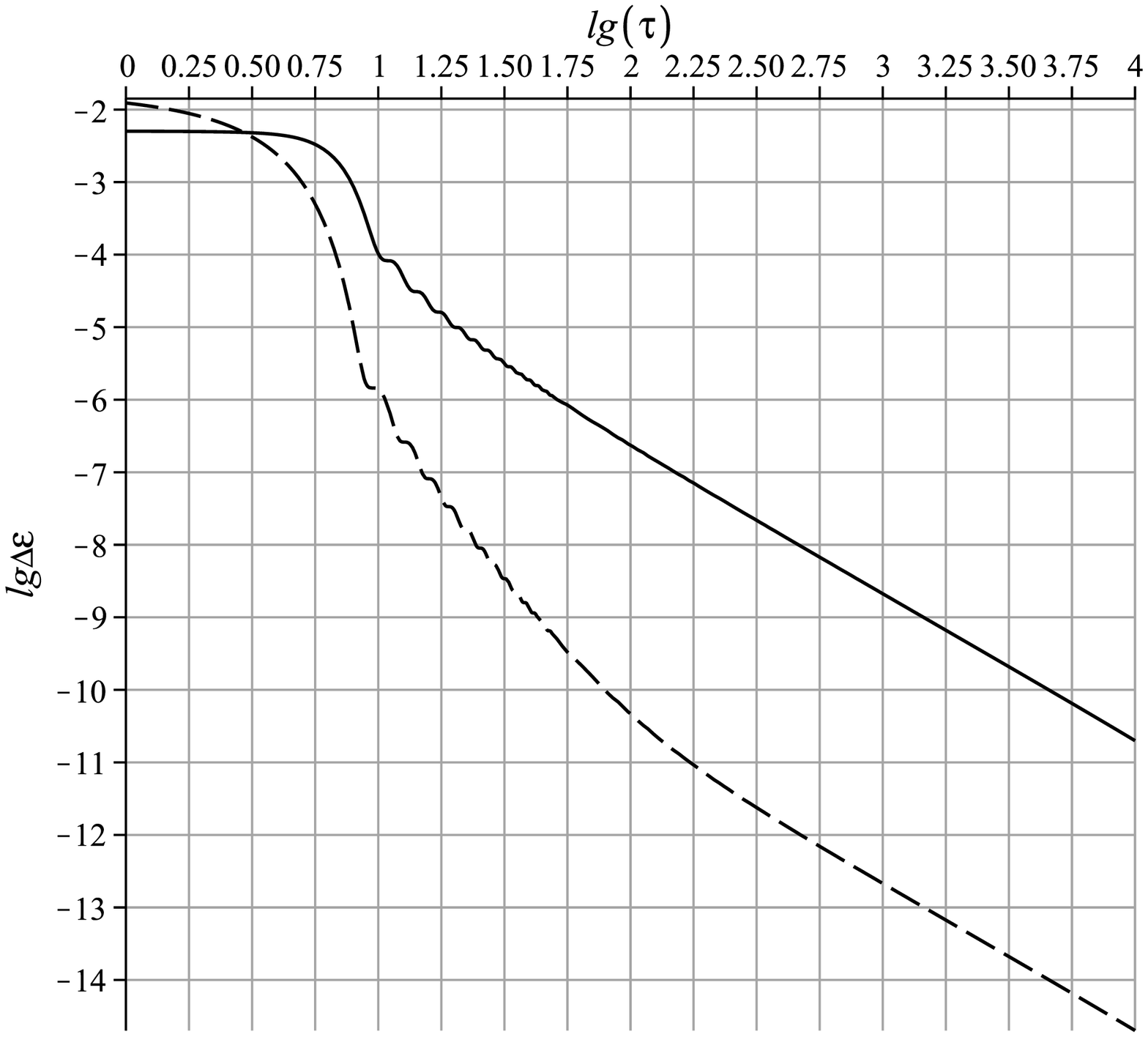}{Evolution of the energy density $\varepsilon _{0} (\tau )$  (dotted line) and mean-square correction to it,  $\Delta \varepsilon (\tau )$,  (solid line).\label{img-8}}
\Fig{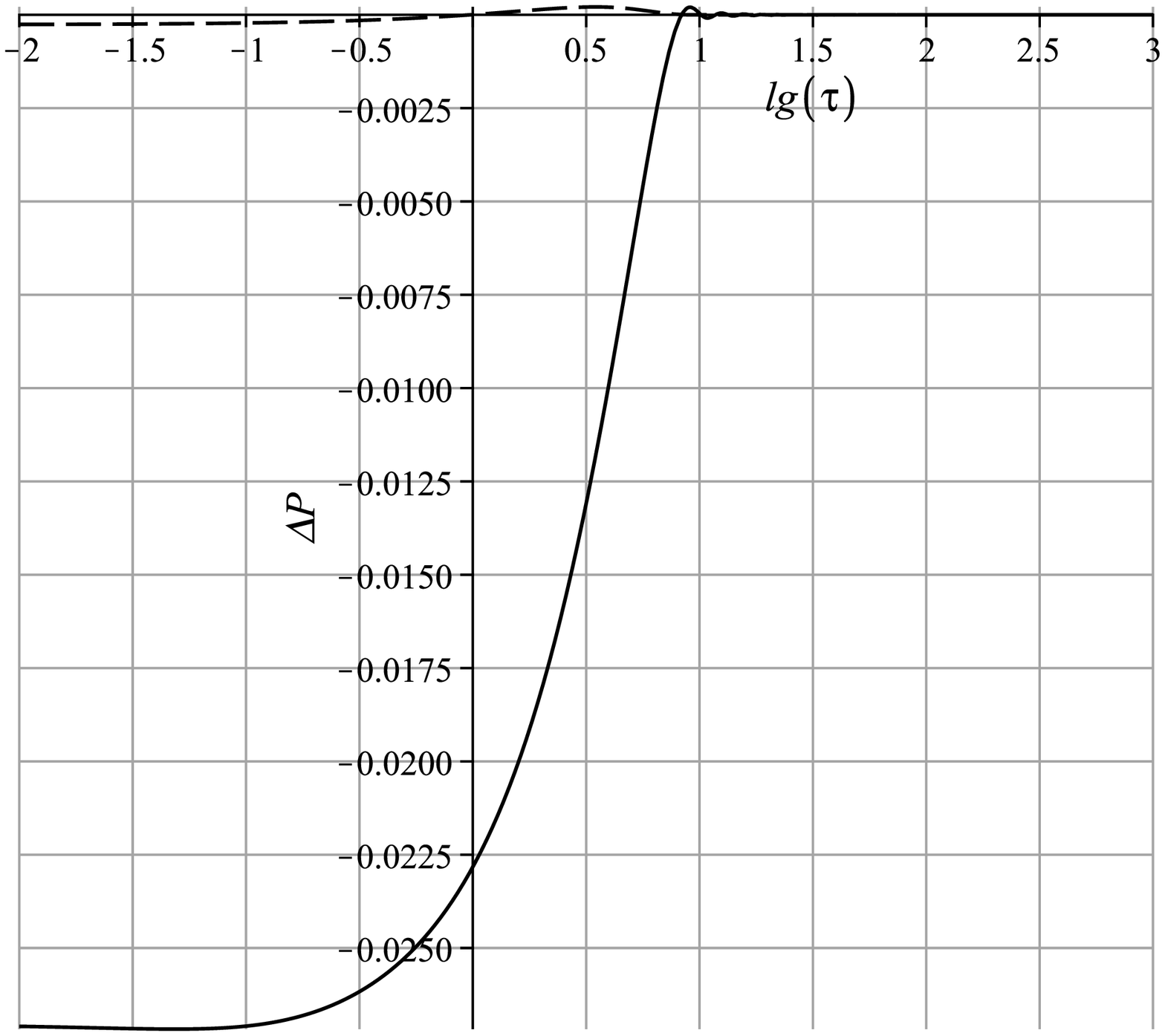}{Evolution of the scalar field's pressure, $p_{0} (\tau )$, (dotted line) and mean-square correction to it, $\Delta p(\tau )$, (solid line).\label{img-9}}

\section{The Discussion of the Results}
Let us proceed to analysis of the numerical si\-mu\-la\-tion results. Let us recall that all calculations are done in Compton time scale, that is why oscillations of the dynamic variables are strictly of mic\-ro\-s\-copic character. i.e. are nonobservable values in macroscopic observations. The analysis of graphs represented on Fig.8 leads, firstly, to an unexpected conclusion: \textit{on the later evolution times the mean - square correction to scalar field's energy density, stipulated by mic\-ro\-s\-copic oscillations of this field, multiple (up to $10^{4} $ times!)exceeds the co\-n\-t\-ri\-bution to energy density, calculate with respect to averages of the scalar field and its derivative}. This means that at large times of the early Universe the co\-n\-t\-ri\-bution of the energy of scalar field's microscopic oscillations is an overwhelming. However, it starts to prevail since times of the order of$10t_{c} $.

Second, our attention is attracted by the fact that according to (\ref{GrindEQ__16_}) there should fulfill the next asymptotic relation
\[\overline{(\Delta \Phi ')^{2} }=\overline{(\Delta \Phi )^{2} }\Rightarrow \Delta p\to 0{\rm \; \; }(\tau \to \infty ).                                              \]
This means that the effective microscopic equa\-tion of state of the oscillating part of the scalar field at late stages is the non - relativistic equation of state $p=0$ (see (\ref{GrindEQ__9_})). The results of papers  \cite{Ignat16-1} -- \cite{Ignat17-1}, which revealed the microscopic oscillations of the invariant cosmological acceleration and showed that average acceleration tends to a non - relativistic value, are also aligned and confirm this outcome. Let us note, that an oscillating mode of the invariant cosmological acceleration was also discovered in later works, where the evolution of the cosmological plasma of scalar charged particles \cite{Ignat_Mif} -- \cite{Yu_Ass} was considered. Fig. \ref{pe32} shows that oscillations of the invariant cosmological acceleration at sufficiently great times are strictly limited by  $[-1,+1]$.
\Fig{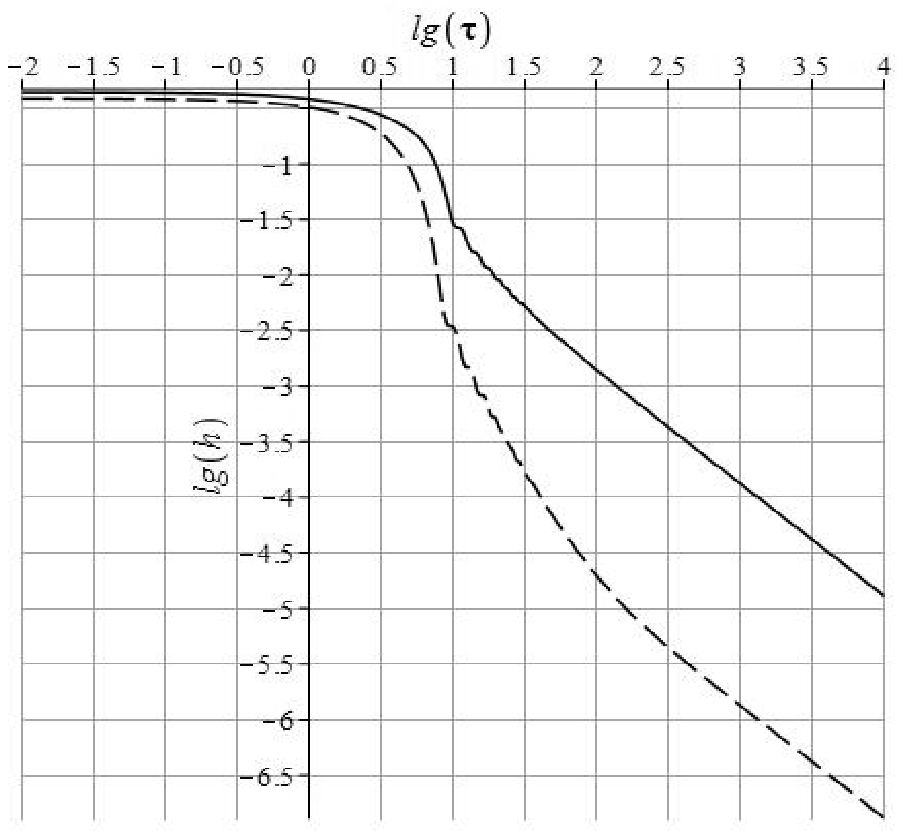}{The evolution of the reduced Hubble constant h=H/m. The dotted line is the Hubble constant, which is calculated relative to average values $\Phi _{0} (\tau )$è$Z_{0} (\tau )$ by formula (\ref{GrindEQ__11_}). The solid line is the Hubble constant, calculated using averaging by formula (\ref{GrindEQ__22_}).\label{img-10} }
\Fig{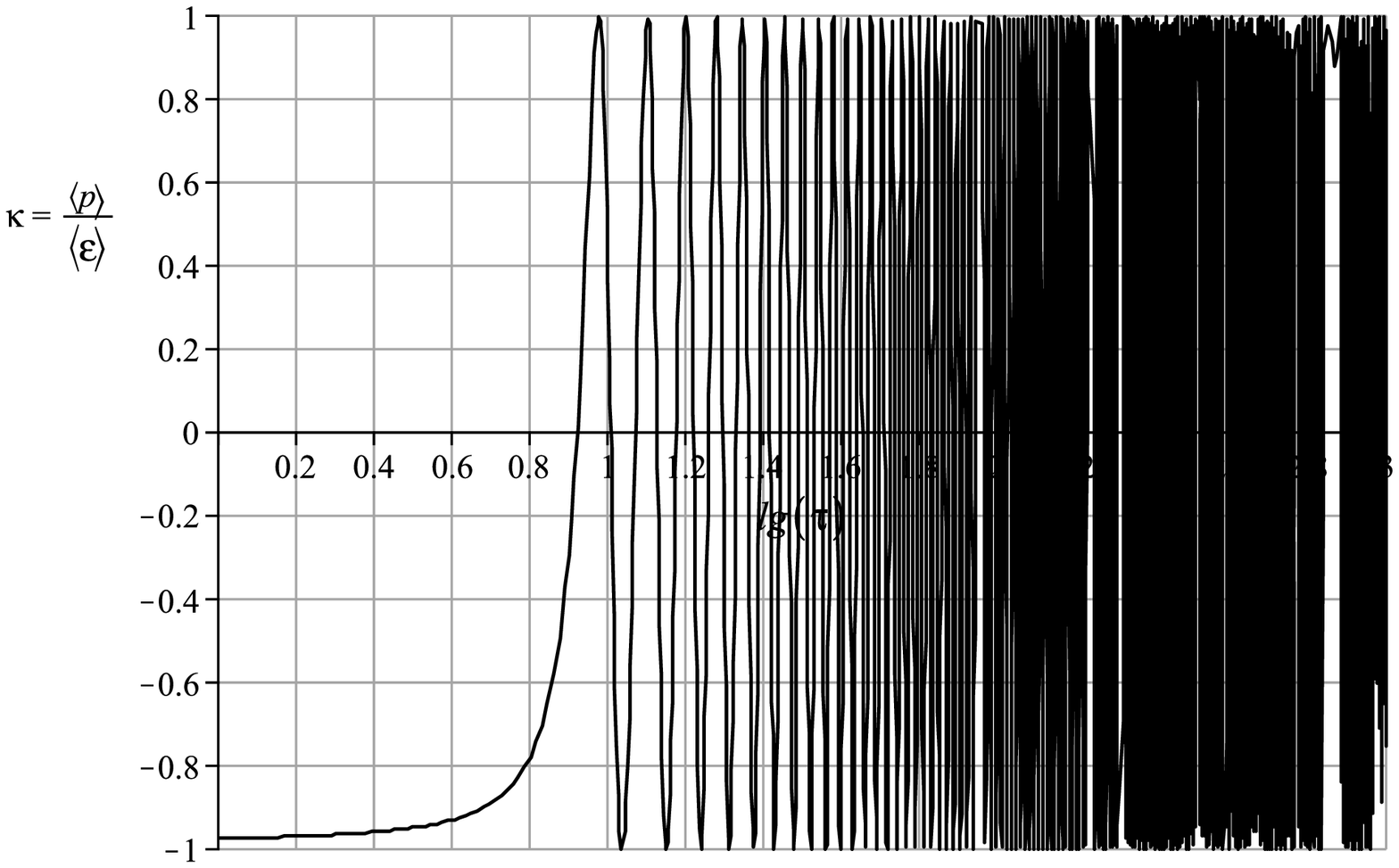}{The evolution of the relation $\overline{p}/\overline{\varepsilon}$. Values of $\lg\tau$ are put on the abscissa axis. The averaging is carried out by interval $16T=32\pi$. \label{pe32}}
 There arises the question whether it is possible to interpret microscopic oscillations of the scalar field in terms of quantum theory as generation of massive bosons with mass $m$? In such a case the value $n={\Delta \varepsilon \mathord{\left/ {\vphantom {\Delta \varepsilon  m}} \right. \kern-\nulldelimiterspace} m} $ could be interpreted as a density of these bosons. Actually, the oscillations by law (\ref{GrindEQ__16_})
\[\Phi \simeq \Phi _{0} (t)e^{itm} ;{\rm \; \; \; }\Phi _{0} (\tau )\simeq \frac{1}{\sqrt{3} \tau } ,\quad (\tau {\rm \gg }1)\]
are very similar to a wave function of non - relativistic bosons $\psi \sim e^{iEt+i\mathbf{pr}} $ with $m$ mass and zero mo\-men\-tum.

Due to the non - relativistic character of mac\-ros\-copic tensor of energy - mo\-men\-tum of the scalar field os\-cil\-la\-tions, their total energy per volumed unit can be defined as
\[\Delta E(\tau)= a^3(\tau)\Delta\varepsilon(\tau) = \mathrm{e}^{3\Lambda_0(\tau)}\Delta\varepsilon(\tau).\]
Fig. \ref{DeltaE} shows the evolution of the total energy of the scalar field's oscillations. The maximum is reached in a scale of order of several Compton times.
\Fig{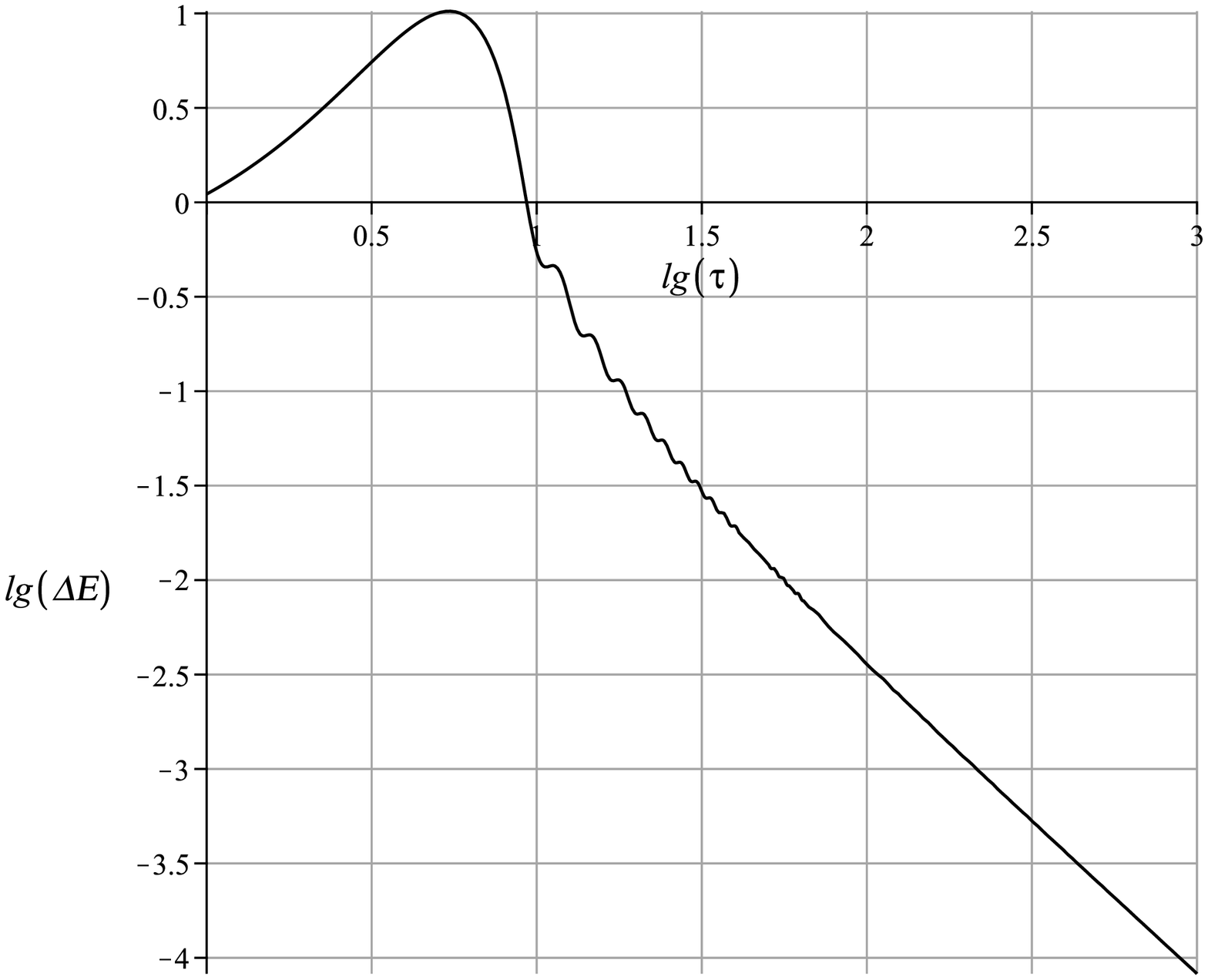}{The Evolution of the Macroscopic Total Energy of the Scalar Field's Oscillations.\label{DeltaE}}

Which cosmological consequences can the factor of energy prevalence of the microscopic oscillations lead to? Since microscopic oscillations are una\-vai\-lable for observations, an observer virtually mea\-sures an average energy of microscopic oscillations by the Hubble constant. An observer can interpret this energy as a contribution from a certain type of non - relativistic dark matter, which practically can represent non - relativistic scalar bosons with the rest mass $m$.

\section*{Acknowledgements}
The Authors express their gratitude to participants of the seminar for relativistic kinetics and cosmology (MW) of the Kazan Federal University for the helpful discussion of the work. Also, the Authors express their gratitude to V. M. Zhuravlev for raising a question on existence of a certain cycle in the phase space $\{\Phi,\dot{\Phi}\}$, which fruitfully influenced on the research of microscopic oscillations of the scalar field.


\begin{thebibliography}{99}
%
\bibitem{Gorb}
%
D.S. Gorbunov and V.A. Rubakov, Introduction to the Theory of the Early Universe: Cosmological
Perturbations and Inflationary Theory. Singapore: World Scientific (2011).
%
\bibitem{Belinsky}
V.A. Belinskii, L.P. Grishchuk, Ya.B. Zel'dovich, I.M. Khalatnikov, Sov. Phys. JETP 62(2), 195-203 (1985).
%
\bibitem{Zeld}
A. D. Dolgov, Ya. B. Zeldovich and M. V. Sazhin. Cosmology of early Univesary, Moskow, Moskow University, 1988.
%
\bibitem{Zhur_01}
V. M. Zhuravlev, JETP, 2001 V. 20., No 5, 1042.
%
\bibitem{Bron}
K. A. Bronnikov and S.G. Rubin, Lection on Gravitation and Cosmology, 2008, Moskow, MIPI.
%
\bibitem{Mex1}
Urena-Lopez L.A., Reyes-Ibarra M.J. arXiv:0709.3996v2 [astro-ph]. 2009.
URL: {\tt https://arxiv.org/pdf/0709.3996.pdf}
%
\bibitem{Zhur}
Zhuravlev V.M., Podymova T.V., Pereskokov E.A.  Grav. and Cosmol. 2011. Vol. 17. ¹ 2.  P. 101.
%
\bibitem{Mex2}
Urena-Lopez L.A.   arXiv:1108.4712v2 [astro-ph.CO]. 2012.
%
\bibitem{Ignat16-1}
Yu. G. Ignat'ev, Space, Time and Fund. Interact., 2016, Issue 3 (16), 37; arXiv:1609.00745 [gr-qc].
%
\bibitem{Ignat16-2}
Yu. G. Ignat'ev, Space, Time and Fund. Interact., 2016, Issue 3 (16), 16; arXiv:1609.08851[gr-qc].
%
\bibitem{Ignat17-1}
Yu. G. Ignat'ev. Grav. and Cosmol. 2017, Vol. 23, No. 2, pp. 131.
\bibitem{Ignat_Agafon17}
Yu. G. Ignat'ev and  A. A. Agathonov, Space, Time and Fund. Interact., 2017 Issue 1(18), 46.
%
\bibitem{Ignat_Sam}
Yu. G. Ignat'ev and  A. R. Samigullina, 2016, Issue 4 (17), 147.
%
\bibitem{Ignat_Mif}
Yu.G. Ignatyev and R.F. Miftakhov.	Grav. and Cosmol. 2011, Vol. 17, 190.
%
\bibitem{Yu_Ass}
Yurii Ignat'ev, Alexander Agathonov, Mikhail Mikhailov and Dmitry Ignatyev.	Astrophys Space Sci (2015) 357:61.
%
\end{thebibliography}
\end{document}